\documentclass[12pt,letterpaper]{article}
\usepackage{jheppub}                 

\usepackage{color}
\usepackage[usenames,dvipsnames]{xcolor}

\usepackage{epsfig,multirow,subfigure} 
\usepackage{amscd}
\usepackage[matrix,arrow,curve]{xy}
\usepackage{verbatim}
\usepackage{latexsym}
\usepackage{amsfonts,amsthm,amsmath}

\newcommand{\be}{\begin{equation}}
\newcommand{\ee}{\end{equation}}
\newcommand{\bea}{\begin{eqnarray}}
\newcommand{\eea}{\end{eqnarray}}

\subheader{\begin{flushright}
UCSD-PTH-12-16\\ 
IPPP/13/25\\
DCPT/13/50
\end{flushright}}
\title{\centering A zig-zag index}
\author[a]{P.~Agarwal,}
\author[b]{A.~Amariti.}
\author[c]{A.~Mariotti.}

\affiliation[a]{Department of Physics, University of California, \\
San Diego La Jolla, CA 92093-0354, USA}
\affiliation[b]{Laboratoire de Physique 
Th\'eorique de l'\'Ecole Normale Sup\'erieure  \\
and  Institut de Physique Th\'eorique Philippe Meyer\\
24 Rue Lhomond, Paris 75005, France}
\affiliation[c]{Institute for Particle Physics Phenomenology,\\
Department of Physics, Durham University, DH1 3LE, United Kingdom}

\emailAdd{pagarwal@physics.ucsd.edu}
\emailAdd{amariti@lpt.ens.fr}
\emailAdd{alberto.mariotti@durham.ac.uk}

\abstract{ We study the large $N$ superconformal index of quiver gauge
  theories describing the worldvolume of D3 branes probing toric
  Calabi Yau singularities.  The index has been previously noticed to
  factorize over the set of the extremal BPS mesonic operators of the
  gauge theory.  We review this factorization and reformulate it in
  terms of zig-zag paths in the dimer model associated to the quiver.
  By using this reformulation, we argue that the factorization is
  valid not only for the exact $R$-charge but for every set of
  $R_{trial}$ respecting the marginality constraints.  Moreover, we
  show the factorization of the index also in theories with orbifold
  singularities, previously not investigated.  We conclude by
  providing an expression for the index in terms of the toric data of
  the dual geometry.  }

\begin{document}

\maketitle

\section{Introduction}
\label{Sec:Introduction}

The superconformal index (SCI) of four dimensional superconformal
field theories \cite{Romelsberger:2005eg,Kinney:2005ej} 
is the supersymmetric partition function of the theory defined
on the euclidean space $S^3 \times
S^1$.  Alternatively, it can be defined as a weighted (over the
fermion number) sum of the states of the theory, where the
contribution of the long multiplets vanishes.  The index
counts the short BPS multiplets and it is invariant
under marginal deformations of the theory.  It has been extensively
studied in the recent years, especially to check field theory dualities and
the AdS/CFT correspondence
\cite{Romelsberger:2005eg,Kinney:2005ej,Romelsberger:2007ec,
  Dolan:2008qi,Spiridonov:2008zr,Spiridonov:2009za,
  Gadde:2010en,Eager:2012hx}.

There are many prescriptions for obtaining the functional form of the
index \cite{Romelsberger:2005eg,Kinney:2005ej,Sen:1985dc,Sen:1985ph,
  Romelsberger:2007ec,Nawata:2011un,Agarwal:2012hs}.  In the large $N$
limit, the computation of the index simplifies and in some cases it
can be carried over with matrix model techniques.

In this paper, we focus on a large class of superconformal gauge
theories, namely the quiver gauge theories arising as the world volume
of D3 branes probing a toric CY$_3$ singularity.  It has been shown
that the large $N$ index for such theories can be computed, matches
with the dual description, and that it usually factorizes on a specific
subset of operators, the so called extremal BPS mesons, corresponding
to the edges of the dual cone of the toric fan.

This factorization was first observed in \cite{Gadde:2010en} for the
SCI of of the $Y^{pq}$ families \cite{Benvenuti:2004dy} of quiver
gauge theories.  By fixing the value of the superconformal $R$-charge
imposed by $a$-maximization the authors computed the index in the
$Y^{p0}$ and $Y^{pp}$ theories and guessed a general behavior for the
$Y^{pq}$ case. A proof for the conjecture was later provided in
\cite{Eager:2012hx}, where the authors explained the factorization of
the index from the properties of the toric geometry, for the case of
smooth CY$_{3}$'s.

In this paper we show that the factorization property of the SCI for
toric quiver gauge theories is more general.  First, we observe that
the index factorizes without fixing the exact superconformal
$R$-charge, but just by requiring that the NSVZ beta functions vanish
and the superpotential is marginal \footnote{With a slight abuse of
  notation we keep on referring to this supersymmetric partition
  function on $S^3 \times S^1$ as the superconformal index also in
  this case.}.  Second, we show that the factorization holds also in
gauge theories dual to geometries with additional singularities.

For this purpose, we reformulate the factorization of the SCI on 
extremal BPS  mesons as a factorization of the SCI over a set of paths in
the brane tiling. These paths are called zig-zag paths because they
turn maximally left (right) at the black (white) nodes of the
bipartite tiling.  We conjecture a general factorized formula for
the SCI in terms of the zig-zag paths, as a function of a trial 
$R$-charge.  This expression continues to be well defined in the case of
quiver gauge theories dual to geometries with orbifold singularities.

We check the validity of our formula and the factorization of the 
SCI index over the zig-zag paths in various examples,
including infinite families of orbifold singularities.  Moreover, we
verify the invariance of our formula under Seiberg duality.
As a byproduct, the factorization over the zig-zag path allows us
to express the SCI directly in terms of the CY geometry and the toric
data.

The paper is organized as follows. In section \ref{Sec:SCI} we review
the relevant aspects of D3 branes at toric CY$_3$ singularities and of
the large $N$ calculation of the superconformal index.  In section
\ref{Sec:Fac} we explain the factorization of the index over the
extremal BPS mesons as discovered in \cite{Eager:2012hx}.  In
\ref{SubSec:ExtremalEzz} we give the prescription to relate the 
$R$-charges of the extremal BPS mesons to the ones of the zig-zag paths and
we re-formulate the factorization in terms of these paths.  In section
\ref{SubSec:ex} we study the factorization over the zig-zag paths, in
some simple examples, for general values of the trial $R$-charges that
satisfy the constraints imposed by marginality.
In section \ref{Sec:Sing} we prove the factorization in the infinite
families of $L^{aba}$ non-chiral singularities.  In section
\ref{Sec:SD} we show that our formula is preserved by Seiberg
duality. In section \ref{Sec:Glo} we show the role of the global, non
anomalous and non $R$-symmetries in the factorization. In section
\ref{Sec:Geo} we translate the index from the zig-zag paths to their
geometric counterpart.  We conclude in \ref{Sec:Conc} with some open
problems.  In appendix \ref{Ypqfam} we compare the zig-zag
factorization with the one discovered in \cite{Gadde:2010en} for the
whole $Y^{pq}$ family

\section{Review: SCI and toric quivers}
\label{Sec:SCI}
\subsection{D3 branes on toric CY$_3$}
\label{SubSec:D3CY3}
In this section we  review some aspects of the world-volume theory
describing D3 branes probing a toric CY$_3$ singularity, that will be useful for
the rest of the paper (see \cite{Kennaway:2007tq} and references therein for a 
comprehensive review).

We start by the definition of a quiver gauge theory.  A quiver is a
graph made of vertices with directed edges connecting them.  The
vertices represent the $SU(N)$ gauge groups and the edges represent
bifundamental or adjoint matter fields.  The direction of the arrow
of an edge is associated to the representation of the corresponding
matter field under the gauge groups.

Since we study SCFTs there are two classes of constraints imposed by
superconformality, both associated to the vanishing of the beta
functions. 

The first constraint comes from requiring the vanishing of the NSVZ
beta function for each gauge group. This corresponds to the
requirement of the existence of a non anomalous $R$-symmetry in the
SCFT and hence becomes a constraint on the $R$-charges.  At the $k$-th
node of the quiver we have
\begin{equation}
\sum_{i=1}^{n_{k}}(r_i-1)+2=0
\end{equation}
where the sum is over all the $n_{k}$
bifundamentals charged under the $k$-th gauge group.  
The second constraint comes from imposing the
marginality of the superpotential terms. 

These two constraints restrict the possible $R$-charge
assignments of the superconformal field theory to a 
subset named $R_{trial}$. The extra freedom
is fixed through a-maximization  \cite{Intriligator:2003jj}, 
that gives  eventually the exact $R$-charge.
In the following
we refer to the case
where $R$ is exact as the \emph{on-shell} case, while the case obtained
by just imposing the marginality constraints is referred as 
the \emph{off-shell} case.

Note that in general the superpotential cannot be read from the
quiver, but in the case of toric CY it is possible thanks to the
notion of planar quiver.  Toric quiver gauge theories have the
property that each field appears linearly in the superpotential and in
precisely two terms with opposite signs. It can be shown that we
can exploit this structure of superpotential terms to transmute the
quiver into a planar quiver embedded in $ T^2$. The planar quiver is
thus a periodic quiver built from the original one by separating all
the possible multiple arrows connecting the nodes such that
corresponding to each superpotential term there is a plaquette whose
boundaries are given by the arrows,  the bifundamental
fields appearing in that superpotential term. Plaquettes representing
superpotential terms with a common bifundamental are  glued
together along the corresponding edge. The sign of a superpotential
term  corresponds to orientation of its plaquette.

Moreover, 
it is possible to define a set of paths on the planar quiver called zig-zag
paths.  They are loops on the torus defining the planar quiver.  These
loops are composed by the arrows. These arrows are chosen such that
if a path turns mostly left at one node it turns
mostly right at the next one.
This notion is not illuminating on the quiver but it becomes more
important in the description of the moduli space on the dual graph,
called the bipartite tiling or the dimer model.

The dimer model is built from the planar quiver 
by reversing the role of the faces and of the
vertices. The superpotential terms become the
vertices of the tiling, and the orientation is absorbed in the color
(black or white), i.e. the tiling is bipartite. The edges are mapped
to dual edges, and the orientation is lost (all the information is in
the vertices). The faces represent the gauge groups.

The zig-zag paths are oriented closed loops on the tiling with non
trivial homology along the $T^2$.  Every
node of the tiling is surrounded by a closed loop made out of the 
zig-zag paths, and the orientation of the loops determines the color of
the vertices, consistently with the bipartite structure of the tiling.

On the bipartite tiling there are sets of edges, called perfect
matchings, that connect black and white nodes, such that every node is
covered by exactly one edge.
As already mentioned, the
tiling is defined on the torus, that possesses two winding cycles
$\gamma_\omega$ and $\gamma_z$. 
An intersection number with the homology classes $(1,0)$ and
$(0,1)$ of the two winding cycles is associated to each perfect maching.

A monomial in $z^m \omega^n$ is associated to each perfect matching,
where $m$ and $n$ represent the intersection number of the perfect
matching with the cycles $\gamma_\omega$ and $\gamma_z$.  A polynomial
that counts the perfect matchings in the brane tiling is obtained by
summing over these monomials

The convex hull of the exponents of this polynomial is a polyhedral on
$\mathbf{Z}^2$, the toric diagram. This rational polyhedral
encodes the informations of the moduli space of the D3 probing the
toric CY$_3$.

\subsection{Large N index in toric quivers}
\label{SubSec:LargeN}
The superconformal index for a four dimensional $\mathcal{N}=1$
field theory is defined as
\begin{equation}
I = Tr (-1)^F e^{-\beta \Xi} t^{R-2J_3} y^{2 \tilde J_3} \prod \mu_i^{q_i} 
\end{equation}
where $\Xi = \{ Q_1, Q_1^\dagger\}$ represents the superconformal
algebra on $S^3 \times S^1$.  The index gets contributions only from
the states with $\Xi=0$ and hence it is independent from $\beta$.

The chemical potentials $t$, $y$ and $\mu$ are associated to the
abelian symmetries of the theory that commute with $Q_1$
and their charges are the exponents, $R$ is the
$R$-symmetry, $J_3$ and $\tilde J_3$ are the Cartan of the
$SU(2)_L\times SU(2)_R \in SO(4,2)$ and $q_i$ are the charges of the
flavor symmetries.  The single particle index receives contributions
from both the chiral and the vector multiplet.  In the first case we
have
\begin{equation}
I_{s.p}(\phi) = \frac{t^{r_\phi}}{(1-t y )(1-t/y)}
\quad,\quad
I_{s.p}(\psi^{\dagger}) = -\frac{t^{2-r_\phi}}{(1-t y )(1-t/y)}
\end{equation}
where both $\phi$ and $\psi$ belong to the chiral multiplet $\Phi$.
The contribution of the vector multiplet is
\begin{equation}
I_{s.p.}({\bf V}) = \frac{2 t^2-t(1+1/y)}{(1-t y)(1-t/y) }
\end{equation}

In the case of quiver gauge theories there are only two possible
representations, bifundamental and adjoint.  A bifundamental
superfield $X_{ij}$ contains a scalar in the fundamental for the
$i$-th group and in the antifundamental for the $j$-th group.  The
fermion $\psi^{\dagger}$ is in the opposite representation.

The single particle index $I(t,y,\chi)$ associated to the quiver is
the sum of the contributions of the vector multiples and the
bifundamental multiplets in the quiver.  At each node $i$ there is a
contribution  $I_{s.p.} (V_i) \chi_i^{adj}$, where $\chi_i^{adj}$ is the
character of the adjoint representation of the $i$-th gauge
group. For every bifundamental $\Phi_{ij}$ there is a contribution
\begin{equation}
\label{sbp}
I_{i,j}(t,y,\chi)=
I_{s.p}(\phi_{ij})\chi_{i}\bar{\chi_{j}}+I_{s.p.}(\psi_{ji}^{\dagger})
\bar \chi_{i}{\chi_{j}}
\end{equation}
where the $\chi_{i}$ and $\bar \chi_{i}$ are the characters of the
fundamental and antifundamental representation associated to the
$SU(N_i)$-th gauge group.  If the matter field is a bifundamental the
product $\chi_{i}\bar{\chi_{j}}$ in (\ref{sbp}) must be substituted
with $\chi_i^{adj}$.

The single trace index is 
obtained by taking the plethystic exponential \cite{Benvenuti:2006qr}.
In order to single out
contributions from gauge-invariant states, we also need to integrate
over the gauge measure. In formulae 
\begin{equation}
I_{m.t.}(x) = \int \prod_{i=1}^{G}[d\alpha_i] PE[I(t,y,\chi(\alpha_i))]
\end{equation}
where the $\alpha_i$ are the Cartan of the $i$-th gauge group.
By taking the large $N$ limit this becomes a Gaussian integral
and the index is
\begin{equation}
I_{m.t.}(t,y)=
\prod_k
\frac
{e^{\frac{1}{k} \text{Tr}\,i(t^k,y^k)}}{\det(1-i(t^k,y^k))}
\end{equation}
where 
\begin{equation}
1-i(t,y) = \frac{1-m(t)+t^2 m^T(t^{-1})-t^2}{(1-t y)(1-t/y)}\equiv
\frac{M(t)}{(1-t y)(1-t/y)}
\quad
\text{with}
\quad
m_{ij}(t) = \sum_{e:i\rightarrow j} t^{R(e)} 
\end{equation}
The matrix $m(t)$ represents the adjacency matrix weighted by the
$R$-charge. For every edge $e$, connecting the $i$-th node
to the $j$-th one in the quiver, the matrix picks up a contribution $t^{R(e)}$.
The index can be further simplified 
and it becomes
\begin{equation} \label{st}
I_{s.t.} (t,y) = -\sum_{k=1}^{\infty}
\frac{\varphi(k)}{k} \log\det M(t^k)
-\text{Tr}
\left(\frac{m(t)-t^2 m(t^{-1})}{(1-t y)(1-t/y)}\right)
\end{equation}
where $\varphi$ is the Euler-phi function.  Observe that the second
term in (\ref{st}) vanishes in absence of adjoint matter because $m(t)$
becomes traceless.

\section{Factorization of the SCI}
\label{Sec:Fac}
\subsection{SCI over the extremal BPS mesons}
\label{SubSec:extrBPS}

The factorization of the index was first observed in
\cite{Gadde:2010en} and then  proven in \cite{Eager:2012hx} for toric CY$_3$
without additional singularities away from the tip of the cone.

Consider a toric CY$_3$
cone probed by a D3 brane.  This cone is described by the fan
$\mathcal{C}$, a convex polyhedral cone in ${R}^3$.  The BPS mesons
(their vev), up to F-term equivalences, are in 1-1 correspondence with
the semigroup of integer points in $\mathcal{C^*}$,  the dual cone 
of $\mathcal{C}$.  The three integer numbers defining the points in the
dual cone (and equivalently the BPS mesons) are the three $U(1)$
isometries of the CY$_3$ or equivalently the mesonic symmetries of the
field theory ($U(1)_F^2 \times U(1)_R$).  
The points in the dual cone can be divided 
in points on the edges, on the faces and on the internal
of the cone itself.

After this geometrical digression we can now report the result of
\cite{Eager:2012hx} on the factorization of the index.  It states that
the determinant $\det(M(t))$ factorizes over the extremal BPS mesons
\cite{Benvenuti:2005ja}
that are described by the edges of the dual cone $\mathcal{C^*}$
\begin{equation}
\label{resufact}
\det (M(t)) = \prod_{i \in E_{M}}
\left(1-t^{r_i} \mu_1^{F_i} \mu_2^{\widetilde F_i} \right)
\end{equation}
where $E_{M}$ refers to the edges of the dual cone or equivalently to
the extremal BPS mesons.
The charges appearing in \cite{Gadde:2010en,Eager:2012hx} 
are the exact $R$-charge
of the SCFT and the two $U(1)_F$.

There are some interesting questions following from the
factorization.  The first regards the exactness of the $R$-charge. One
may wonder if the exact $R$-charge is a necessary condition for the
factorization of the index, or if it possible to relax this
assumption, just by imposing the marginality constraints (vanishing of
the beta functions),
corresponding to the \emph{off-shell} $R_{trial}$ case defined above.

A second question regards theories with extra
singularities far from the tip of the cone. These theories are
characterized by having extra points on the edges of the toric
diagram. In the dual cone these points are not associated to any edge
but they live on the faces. These theories have not been investigated in
\cite{Eager:2012hx} and one may wonder how the 
factorization formula is modified in these cases.

\subsection{Extremal BPS mesons and zig-zag paths}
\label{SubSec:ExtremalEzz}

In this section we study the two problems discussed above by using the
brane tiling instead of the dual cone.  By starting from the
observation that both the extremal BPS mesons and the zig-zag paths
are in 1-1 correspondence with the primitive vectors of the toric
diagram we give a prescription to extract the charges of the extremal
BPS mesons from the charges of the zig-zag paths.  This allows us to
define a factorization formula for the SCI in terms of the zig-zag paths.

The BPS mesons, not necessarily extremal, are represented on the
tiling as string of operators built by connecting a face with its
image by a path.  These paths have to cross the edges of the tiling by
leaving the nodes of the same color on the same side.  A BPS meson is
the product of the edges crossed by such paths.  Products of operators
with the same homology and the same $R$-charge are $F$-term
equivalent.  There is a set of these BPS mesons that have maximal
$U(1)$-charge (up to a sign) for a given $R$-charge.  These are the
extremal BPS mesons, corresponding to the edges of the dual cone
\cite{Benvenuti:2005ja}.

They can be built (up to degenerations) from the zig-zag paths.  First
we associate an orientation to every zig-zag path such that they leave
a black node on the right and a white node on the left.  For every
black $n$-valent node \footnote{The same correspondence can be
  obtained by using the white nodes crossed by the $i$-th zig-zag
  path} the $i$-th zig-zag path crosses two edges.  The $i$-th extremal BPS
meson is obtained by associating the other $n-2$ edges at every black
node crossed by the $i$-th zig-zag path.

\begin{figure}
\begin{center}
  \includegraphics[width=15cm]{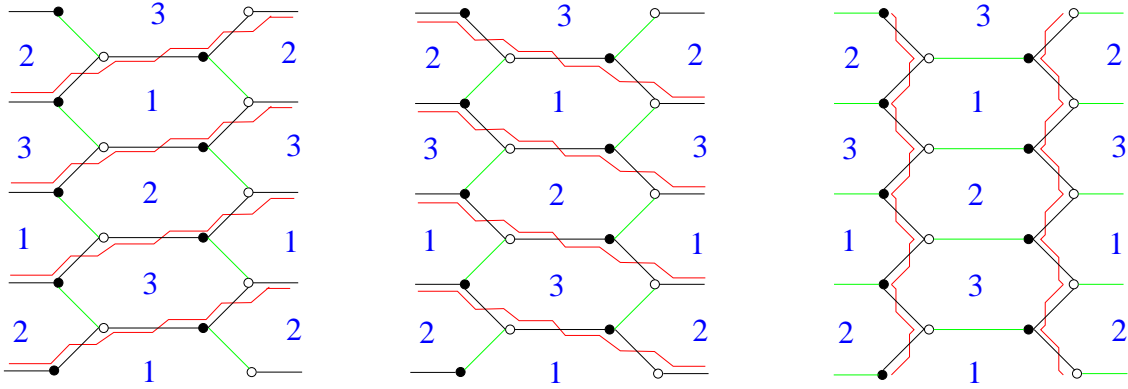}
  \caption{Zig-zag paths and extremal BPS mesons for
    $\mathbb{C}^3/\mathbb{Z}_3$ .}
\label{fig:zigBPS}
\end{center}
\end{figure}

For example in the figure \ref{fig:zigBPS} we highlight in red the
three zig-zag paths of $\mathbb{C}^3/\mathbb{Z}_3$ and in green the
three extremal BPS mesons.
From this definition we obtain a general formula relating the $R$-charges 
of the extremal BPS mesons and the $R$-charges of the zig-zag paths.

At each $n$-valent black node the condition of marginality of the
superpotential implies that
\begin{equation} \label{eq:sumch}
\sum_{j=1}^{n} r_j=2
\end{equation}
where $r_j$ are the charges of the fields 
related to the edges connected with the
black node that we are considering.

Let us suppose that the first two ($j=1,2$) are in the zig-zag paths and
the others in the extremal BPS meson.
By using the previous relation we have that
\begin{equation} \label{eq:rBPS_rZZ}
r_{3}+\dots+r_{n} = 2-r_{1}-r_{2}= (1-r_{1}) +(1-r_{2})
\end{equation}
and we have expressed the $R$-charges of the fields forming the extremal
BPS meson in terms of the $R$-charges of the edges belonging to the
zig-zag path.

By summing over all the black nodes crossed by the zig-zag path we
obtain the $R$-charge of the  extremal BPS  meson associated to the i-th
zig-zag path (denoted with $Z_i$)
\begin{equation} \label{eq:BPS_ZZ_final}
R_{BPS_i} = \sum_{k \in \{Z_i\}}(1-r_{k}^{(i)})
\end{equation}
where $k$ runs over the set of edges $\{Z_i\}$ belonging to the i-th
zig-zag path, and $r_{k}^{(i)}$ is the $R$-charge of the $k$-th field
in the $i$-th zig-zag path.

By using the relation between the $R$-charges of the extremal BPS mesons and of
the zig-zag paths the determinant $\det M(t)$
factorizes over the zig-zag paths as \footnote{In the following we set
  $\mu_1=\mu_2=1$, at the end of the paper we will show how to
  insert these symmetries back in the index.}
\begin{equation}
\label{eq:generalzz}
\det M
=
\prod_{i=1}^Z(1-t^{\sum_{j \in \{Z_i\}}(1-r_{j}^{(i)})})
\end{equation}
where $Z$ is the number of zig-zag paths, and $\{Z_i\}$
and  $r_{j}^{(i)}$ are defined as above.

We conjecture (\ref{eq:generalzz})
to be valid also \emph{off-shell} and in the singular cases.
In the rest of the paper
we study the validity of this 
formula with many examples and checks.

\section{Examples}
\label{SubSec:ex}
In this section we study the two simplest examples of
quiver gauge theories described by a bipartite graph
and associated to a toric CY$_3$ singularity.
They are the $\mathcal{N}=4$ SYM and the conifold.

In both cases we explicitly show how the Gaussian
integral obtained in the large $N$ limit factorizes
over the zig-zag paths \emph{off-shell}.

\subsection{N=4}
\label{SubSubSec:ex1}

We start by considering the $\mathcal{N}=4$ SYM.  We study this theory
as an $\mathcal{N}=1$ theory.  In $\mathcal{N}=1$ notations there is
an $SU(N)$ gauge group and three adjoint fields, that we call $X_1$,
$X_2$ and $X_3$.  The interaction is $W=X_1[X_2,X_3]$ which imposes
$r_{X_1}+r_{X_2}+r_{X_3}=2$.
The three zig-zag paths correspond to the three products of fields
\be
zz_1=X_1 X_2 \quad, \quad zz_2 = X_2 X_3 \quad \quad  zz_3=X_3 X_1 
\ee

In this theory the determinant at large $N$ (\ref{resufact}) is given by
\begin{equation}
\label{detN4}
\det(M(t)) = 1-t^2 + \sum_{i=1}^{3} t^{r_i}+\sum_{i=1}^{3}t^{2-r_i}
\end{equation}
We now show that this determinant factorize in a product over the
zig-zag path as claimed in (\ref{eq:generalzz}), by manipulating each
term in expression (\ref{detN4}).

The term $t^2$ generically corresponds to $t^{2 n_G}$, where $n_G$ is
the number of gauge groups in the quiver, and it can be re-written
from the relation in the dimer as
\begin{equation}
n_{Faces}+n_{Points}-n_{Edges}=0
\rightarrow 
2n_{fields}-2n_W =2n_G
\end{equation}
By imposing the superpotential constraint we have
\begin{equation}
\sum_{i=1}^{Z} \sum_{j\in\{Z_i\}}(1-r_{j}^{(i)})=2 n_G
\end{equation}
In this case we have $t^2 \rightarrow t^{6-2(r_1+r_2+r_3)}$.

The term $\sum t^{r_i}$ can be re-written by using the 
constraints from the superpotential
and it becomes $\sum_{i<j} t^{2-r_i-r_j}$.
In the same way the last term becomes
\begin{equation}
2-r_i =  r_j +r_k= (2-r_i-r_k)+(2-r_j-r_i)  
\end{equation}
By putting everything together  the final formula is 
\begin{equation}
\det(M(t))=
(1-t^{2-r_1-r_2})(1-t^{2-r_1-r_3})(1-t^{2-r_2-r_3})
\end{equation}
which corresponds to the expression (\ref{eq:generalzz}), factorized
over the three zig-zag paths.

\subsection{Conifold}
\label{SubSubSec:ex2}

The second example 
is the worldvolume theory of a stack of $N$ D3 branes
probing the conifold.
This is represented by a quiver gauge theory 
with two gauge groups $SU(N)_1\times SU(N)_2$
and two pairs of bifundamental-antibifundamental $(a_i,b_i)$
connecting them.
The superpotential is $W = \epsilon_{ij}\epsilon_{lk} a_i b_l a_j b_k$
that imposes 
\begin{equation}
\label{larel}
r_{a_1}+r_{a_2}+r_{b_1}+r_{b_2}=2
\end{equation}
At large $N$ the determinant of $M(t)$ is
\begin{equation}\label{sumT11}
1-\sum_{i,j}t^{r_{a_i}+r_{b_j}}+2t^2+\sum_{i \neq j}\left(t^{2-a_i+a_j}+t^{2-b_i+b_j}\right)
-\sum_{i,j}t^{4-a_i-b_j}+t^4
\end{equation}
we can reorganize the sum as a sum over the zig-zag     
paths as follows. There are
four zig-zag paths parameterized by
\begin{equation}
zz_1=a_1b_1\quad,\quad
zz_2=a_2b_1\quad,\quad
zz_3=a_1b_2\quad,\quad
zz_3=a_2b_2
\end{equation}
We keep fixed the first term in the sum (\ref{sumT11}).
The second one becomes
\begin{equation}
\sum_{i,j}t^{r_{a_i}+r_{b_j}} \rightarrow \sum_{i,j}t^{2-r_{a_j}-r_{b_i}}
= \sum_{i=1}^{Z}t^{\sum_{j \in \{Z_i\}}(1-r_{j}^{(i)})}
\end{equation}
the third and the fourth terms can be written together 
and thanks to the relation (\ref{larel}) we have
\begin{eqnarray}
&&
2t^2+\sum_{i \neq j}\left(t^{2-a_i+a_j}+t^{2-b_i+b_j}\right)
\rightarrow
%
\sum_{i=1}^{Z}\sum_{j=i+1}^{Z}
t^{\sum_{k\in\{Z_i\}}(1-r_k^{(i)})+\sum_{l \in \{Z_j\}}(1-r_l^{(j)})}
\nonumber \\
\end{eqnarray}
Also in the fifth term of (\ref{sumT11}) we can insert the
relation (\ref{larel}) and obtain
\begin{equation}
\sum_{i,j}t^{4-a_i-b_j} \rightarrow
\sum_{i=1}^{Z}\sum_{j=i+1}^{Z}\sum_{k=j+1}^{Z}
t^{\sum_{l \in\{Z_i\}}(1-r_l^{(i)})+\sum_{m \in \{Z_j\}}(1-r_m^{(j)})+\sum_{n\in\{Z_k\}}(1-r_n^{(k)})}
\end{equation}
The last term is obtained as already explained 
in the $\mathcal{N}=4$ case.  Finally, by collecting all the terms, we
have
\begin{equation}
\det(M(t))=
(1-t^{2-r_{a_1}-r_{b_1}})
(1-t^{2-r_{a_1}-r_{b_2}})
(1-t^{2-r_{a_2}-r_{b_1}})
(1-t^{2-r_{a_2}-r_{b_2}})
\end{equation}

\section{The singular cases}
\label{Sec:Sing}

The second result that we argue in this paper is that the determinant
of the matrix $M(t)$ arising in the large $N$ calculation of the
superconformal index (see formula (\ref{st})) factorizes over the
zig-zag paths also in the case where new singularities arise far from
the tip of the CY cone.

For example in the $L^{pqr}$ families \cite{
  Benvenuti:2005ja,Franco:2005sm,Butti:2005sw} there are many examples
corresponding to orbifolds. Inside these classes of orbifolds there
are two infinite families, $L^{aaa}$ and $L^{aba}$, associated to
non-chiral theories that can be studied in a unified way.  In this
section we show that $\det (M(t))$ factorizes in both these cases over
the zig-zag paths.  Moreover we study a non chiral case, $L^{264}$
corresponding to the $L^{a,b,\frac{b-a}{2}}$ singular family, and
observe the factorization.

\subsection{The $L^{aaa}$ family}
\label{SubSec:Sing1}

\begin{figure}
\begin{center}
  \includegraphics[width=15cm]{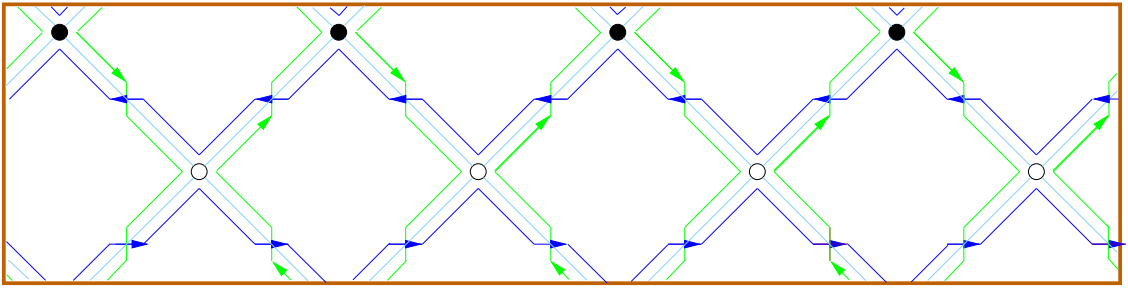}
\caption{Tiling and zig-zag paths for 
 a generic $L^{aaa}$ model.
We grouped the zig-zag paths with homology $(\pm 1,0)$ 
with the green color
while the blue ones have homology  $(0,\pm 1)$.
We distinguished the sign by specifing the orientation with.}
\label{fig:Laaazz}
\end{center}
\end{figure}

In this section we compute the large $N$ index for an infinite class
of theories, the $L^{aaa}$ theories. These theories are vector like
theories with a bifundamental and an antibifundamental connecting the
$i$-th node and the $i+1$-th one. We start by studying the phase without any
adjoint matter field. Subsequently we show that the
factorization of the index over the zig-zag paths is maintained even in
phases that contain the adjoint fields.

\begin{figure}
\begin{center}
  \includegraphics[width=15cm]{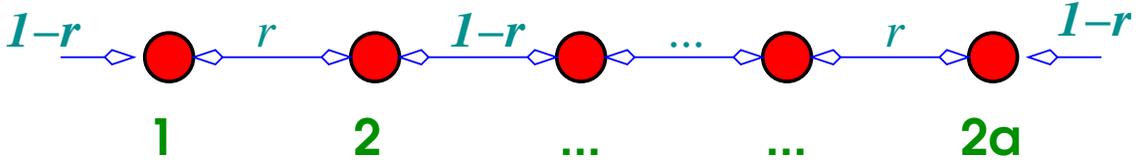}
  \caption{Trial $R$-charge asignation for a generic $L^{aaa}$ model.}
\label{fig:Laaazz2}
\end{center}
\end{figure}

By looking at the tiling there are four classes of zig-zag paths.
The first two classes have homology $(1,0)$ and $(-1,0)$ respectively
and contain $2 a $ fields. By imposing the constraints imposed by the
marginality we have two possible charge assignations, as in figure
\ref{fig:Laaazz2}.
The two zig-zag paths both contribute to the index with a factor
$(1-t^a)$.
There are also other $a$ zig-zag paths with homology $(0,1)$ 
and $a$ with homology $(0,-1)$.
The first class contains only fields with charge $r$, and 
every zig-zag of this kind contributes with a factor $(1-t^{2-2r})$.
In the second case the charge is $1-r$ and the contribution
is $(1-t^{2r})$.
The final contribution to the index is
\begin{equation}
\det M(t) = (1-t^a)^2(1-t^{2-2r})^a(1-t^{2r})^a
\end{equation}
We now give a proof of our claimed factorization.
We start by writing the matrix
\begin{equation}
\label{circulant}
M(t)=\left(
\begin{array}{cccccc}
 a_1 & b_1  & 0 &  \dots &   \dots& c_a \\
 b_1 & a_2  & c_1 &  \dots &\dots  & 0 \\
 0   & c_1 &0 &  & \dots & 0 \\
  \dots  & \dots  &\dots  & \dots & c_{a-1} & 0 \\
 \dots   & \dots  &\dots  & c_{a-1} & a_{2a-1} & b_{a} \\
 c_a & 0   & 0 & 0 & b_{a} & a_{2a} \\
\end{array}
\right)
\end{equation}
where 
\begin{eqnarray}
a_i = (1-t^2)
\quad,\quad
b_{2i} = c_{2i } = (t^{r+1}-t^{1-r})
\quad,\quad
b_{2i+1} = c_{2i+1} = (t^{2-r}-t^{r})
\end{eqnarray}
Since (\ref{circulant}) is a circulant matrix the determinant can be
easily computed. Actually here we use a more complicate technique,
more useful for the $L^{aba}$ case.
The determinant of (\ref{circulant}) can be written 
in an equivalent way by the formula
\begin{equation} \label{Lmat}
\det M(t) = Tr \prod_{i=1}^{2a}
L_i - 2\prod_{i=1}^{a} b_i c_i \quad,\quad
L_{2 j} = \left(
\begin{array}{cc}
 a_j & -b_{j-1}^2 \\
 1 & 0 \\
\end{array}
\right)
\quad 
L_{2 j+1} = \left(
\begin{array}{cc}
 a_j & -c_{j-1}^2 \\
 1 & 0 \\
\end{array}
\right)
\end{equation}
The trace is easily computed by  defining $F = L_{i} L_{i+1}$
and by observing that
\begin{equation}
Tr F^a =Tr \prod_{i=1}^{2a} L_i 
\end{equation}
The trace is  computed from the eigenvalues 
of $F$. We have
\begin{equation}
Tr F^a = Tr
\left(
\begin{array}{cc}
\lambda_1^a  & 0 \\
 0 & \lambda_2^a \\
\end{array}
\right)=
(1 + t^{2 a}) (1-t^{2 -2r})^a (1-t^{2 r})^a
\end{equation}
By adding the extra contribution 
\begin{equation} 
 \prod_{i=1}^{a} b_i\,c_i
=
 t^a (1-t^{2 -2r})^a (1-t^{2 r})^a
\end{equation}
the expected factorization is obtained.

It is interesting to observe the behavior of the index under Seiberg
duality.
As we will show later the factorization of the determinant
is not affected by the duality.
Here the problem is that a duality on the $n$-th node
adds two extra adjoints on the $n\pm 1$-th nodes.
But as we already observed in section \ref{Sec:SCI}
the extra adjoints must be subtracted in the
computation of the index.

While the $\mathcal{N}=1$ vector multiplet
usually cancels the $y$ dependence of the index, 
the presence of the extra adjoints fields
reintroduces this and in principle  
one may expect that the index does 
not match among different phases.
However, this  extra contribution is
\begin{equation}
\frac{1}{(1-t y)(1-t/y)}
\left(
t^{2 r}-t^{2(1-r)}+t^{2(1-r)}-t^{2-2(1-r)}
\right)
\end{equation}
and it vanishes in the dual phase.

\subsection{The $L^{aba}$ family}
\label{SubSec:Sing2}

\begin{figure}
\begin{center}
  \includegraphics[width=15cm]{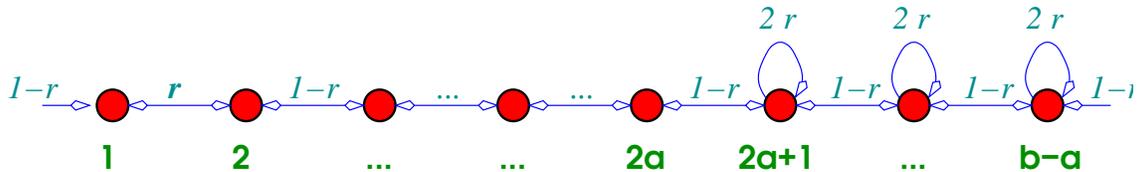}
\caption{Quiver and $R$-charge parameterization for the
$L^{aba}$ theories.  }
\label{fig:Labaq}
\end{center}
\end{figure}
In this section we generalize the case of the $L^{aaa}$ theories
studied above to the whole $L^{aba}$ family.  In this case the
contributions from the extra adjoint matter fields has to be
subtracted, and the index is $y$ dependent.  Nevertheless the
determinant of the matrix $M$ still factorizes over the zig-zag paths.
By parameterizing the fields as in figure \ref{fig:Labaq}there are
four classes of zig-zag paths:
\begin{itemize}
\item $a$ paths formed by the pairs of fields $X_{i,i+1}$ and
  $X_{i+1,i}$ with charge $r$.  They contribute to the index as
  $(1-t^{2(1-r)})^a$.
\item $b$ paths formed by the pairs of fields $X_{i,i+1}$ and
  $X_{i+1,i}$ with charge $1-r$.  They contribute to the index as
  $(1-t^{2 r})^{b}$.
 \item One path formed by all the adjoints and all the fields
   $X_{i,i+1}$. It contributes to the index as $(1-t^{a r+
   b(1-r)})$.
\item One path formed by all the adjoints and all the fields
  $X_{i+1,i}$. It contributes to the index as $(1-t^{a r+
  b(1-r)})$.
\end{itemize}

With the parameterization of the charges in figure \ref{fig:Labaq} the
matrix $M$ is
\begin{equation}
M=
\left(
\begin{array}{cccccccccc}
 a_1 & b_1 & 0 & 0 & 0 & 0 & 0 & 0 & 0 & c_{b} \\
 b_1 & a_2 & c_1 & 0 & 0 & 0 & 0 & 0 & 0 & 0 \\
 0 & c_1 & \dots &  \dots& 0 & 0 & 0 & 0 & 0 & 0 \\
 0 & 0 &  \dots&  \dots& b_a & 0 & 0 & 0 & 0 & 0 \\
 0 & 0 & 0 & b_a & a_{2 a-1} & c_a & 0 & 0 & 0 & 0 \\
 0 & 0 & 0 & 0 & c_a & d_1 & c_{a+1} & 0 & 0 & 0 \\
 0 & 0 & 0 & 0 & 0 & c_{a+1} & d_2 & c_{a+2} & 0 & 0 \\
 0 & 0 & 0 & 0 & 0 & 0 & c_{a+2} &  \dots&  \dots& 0 \\
 0 & 0 & 0 & 0 & 0 & 0 & 0 &  \dots&  \dots& c_{b--1} \\
 c_{b} & 0 & 0 & 0 & 0 & 0 & 0 & 0 & c_{b-1} & d_{b-a} \\
\end{array}
\right)
\end{equation}
where
\begin{eqnarray}
a_i = 1-t^2 ~~~~~~~~& \quad \quad \quad & b_i =
t^{2-r}-t^{r}\nonumber \\ c_i = t^{r-1}-t^{1-r} &
\quad \quad \quad & d_i = 1-t^2-t^{2 r}+t^{2(1-r)}
\end{eqnarray}
As before the determinant of this matrix can be obtained by defining
the two dimensional $L$ matrices (\ref{Lmat}).  The determinant
becomes
\begin{equation}\label{TrLaba}
\det M = \text{Tr} \, \prod_{i=1}^{a+b} L_i +2 \,(-1)^{b+1}\, \prod_{i=1}^a b_i
\prod_{j=1}^{b}c_j
\end{equation}
The first trace can be evaluated by redefining the matrices $L_{i}
L_{i+1}=K$ for $i=1,\dots,2a-1$ and $L_{j}=J$ for $j=2a+1\dots,a+b$.
The trace becomes $Tr K^a J^{b-a}$ where
\begin{eqnarray}\label{KJ}
&& K^a = \frac{\left(\left(1-t^{2 (1-r )}\right)^{a-1}
    \left(1-t^{2 r }\right)^{a-1}\right)}{t^{2 r }} \left(
\begin{array}{cc}
 \left(1-t^{2 r }\right) \left(t^{2 r }-t^{2 (a+1)}\right) &
 -t^2 \left(1-t^{2 a}\right) \left(1-t^{2 r }\right)^2
 \\ \left(1-t^{2 a}\right) t^{2 r } & \left(1-t^{2 r
 }\right) \left(t^{2 (a+r )}-t^2\right) \\
\end{array}
\right)
\nonumber \\
&&
J^{b-a}=
\frac{\left(1-t^{2 r }\right)^{-a+b-1}}{t^{2 r }-t^2} \left(
\begin{array}{cc}
 \left(t^{2 r }-t^{4 r }\right) \left(1-t^{2 (1-r )
   (-a+b+1)}\right) & \left(1-t^{2 r }\right)^2 \left(t^{2
   (1-r ) (b-a)+2}-t^2\right) \\ t^{2 r } \left(1-t^{2
   (1-r ) (b-a)}\right) & \left(t^{2 r }-t^2\right)
 \left(1-t^{2 (1-r ) (-a+b-1)}\right) \\
\end{array}
\right)
\nonumber \\
\end{eqnarray}
After plugging (\ref{KJ}) in (\ref{TrLaba}) we have
\begin{equation}
\det M(t) = \left(1-t^{2-2 r }\right)^a \left(1-t^{2 r
}\right)^{b} \left(1-t^{a r +b(1-r)}\right)^2
\end{equation}
that coincides with the formula computed from the zig-zag paths.

\subsection{A chiral orbifold}
\label{SubSec:Sing3}

\begin{figure}
\begin{center}
\includegraphics[width=15cm]{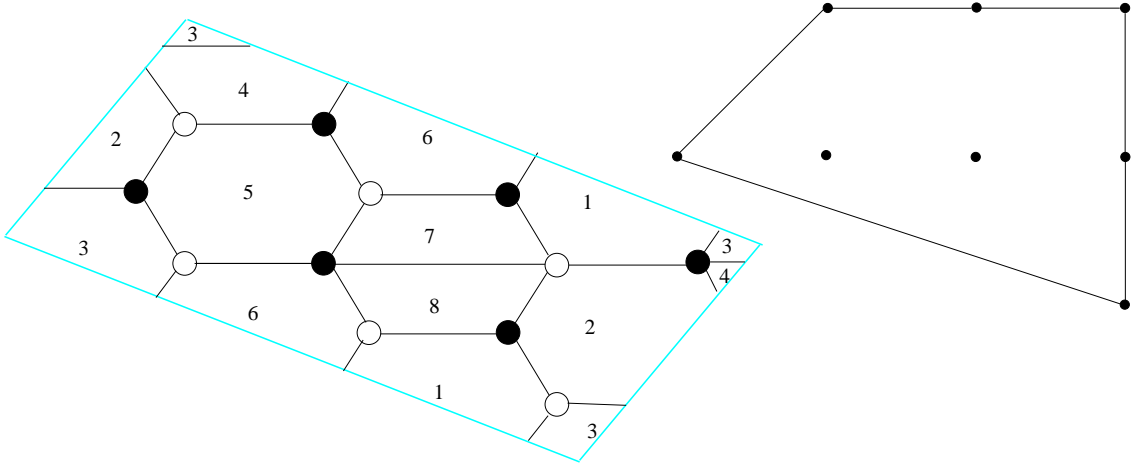}
\caption{Tiling and toric diagram of $L^{264}$}
\label{chiorb}
\end{center}
\end{figure}

We conclude the analysis of the singular cases by studying  
a chiral orbifold of $L^{pqr}$.
This model belongs to an infinite class of chiral orbifolds,
$L^{a,b,\frac{a+b}{2}}$.
We study a single case here,
the $L^{264}$ theory, that is an orbifold of $L^{132}$.
We  show that $\det(M(t))$ factorizes over the zig-zag paths with an
\emph{off shell} $R_{trial}$.
The tiling and the toric diagram are represented in 
(\ref{chiorb})
The matrix $M(t)$ is
\footnotesize{
\begin{equation} M(t) = 
\left(
\begin{array}{cccccccc}
 1-t^2 & -t^{r^X_{1,2}}-t^{r^Y_{1,2}} & t^{2-r^X_{3,1}} & 0 & 0 & -t^{r^X_{1,6}} & t^{2-r^X_{7,1}} & t^{2-r^X_{8,1}}  \\
 t^{2-r^X_{1,2}}+t^{2-r^Y_{1,2}} & 1-t^2 & -t^{r^X_{2,3}} & -t^{r^X_{2,4}} & t^{2-r^X_{5,2}} & 0 & 0 & -t^{r^X_{2,8}}  \\
 -t^{r^X_{3,1}} & t^{2-r^X_{2,3}} & 1-t^2 & t^{2-r^X_{4,3}} & -t^{r^X_{3,5}} & 0 & 0 & 0  \\
 0 & t^{2-r^X_{2,4}} & -t^{r^X_{4,3}} & 1-t^2 & -t^{r^X_{4,5}} & t^{2-r^X_{6,4}} & 0 & 0  \\
 0 & -t^{r^X_{5,2}} & t^{2-r^X_{3,5}} & t^{2-r^X_{4,5}} & 1-t^2 & -t^{r^X_{5,6}}-t^{r^Y_{5,6}} & t^{2-r^X_{7,5}} & 0  \\
 t^{2-r^X_{1,6}} & 0 & 0 & -t^{r^X_{6,4}} & t^{2-r^X_{5,6}}+t^{2-r^Y_{5,6}} & 1-t^2 & -t^{r^X_{6,7}} & -t^{r^X_{6,8}}  \\
 -t^{r^X_{7,1}} & 0 & 0 & 0 & -t^{r^X_{7,5}} & t^{2-r^X_{6,7}} & 1-t^2 & t^{2-r^X_{8,7}}  \\
 -t^{r^X_{8,1}} & t^{2-r^X_{2,8}} & 0 & 0 & 0 & t^{2-r^X_{6,8}} & -t^{r^X_{8,7}} & 1-t^2  \\
\end{array}
\right)
\end{equation}
}
\normalsize
The zig-zag paths are as 
\begin{eqnarray}
&&zz_1 =X_{1,2} \, X_{2,4} \, X_{4,5} \, X_{5,6} \, X_{6,7} \,  X_{7,1} \nonumber \\
&&zz_2 =
X_{1,2}\,X_{1,6}\,X_{2,3}\,X_{2,8}\,X_{3,1}\,X_{4,5}\,X_{5,2}\,X_{6,4}\,X_{6,7}\,X_{7,5}\,X_{8,1}\,Y_{5,6} \nonumber \\
&&zz_3 = X_{2,8}\,X_{3,1}\,X_{4,3}\,X_{5,6}\,X_{6,4}\,X_{7,5}\,X_{8,7}\,Y_{1,2} \nonumber \\
&&zz_4 = X_{2,4}\,X_{3,5}\,X_{4,3}\,X_{5,2} \nonumber \\
&&zz_5 = X_{1,6}\,X_{6,8}\,X_{7,1}\,X_{8,7} \nonumber \\
&&zz_6 = X_{2,3}\,X_{3,5}\,X_{6,8}\,X_{8,1}\,Y_{1,2}\,Y_{5,6}
\end{eqnarray}
After imposing  the NSVZ and the $W$ constraints
we have
\begin{equation}
\det M(t) = \prod_{i=1}^6(1-t^{\sum_{j\in\{Z_i\}}(1-r_{j}^{(i)})})
\end{equation}

\section{Seiberg duality}
\label{Sec:SD}

\begin{figure}
\begin{center}
  \includegraphics[width=12cm]{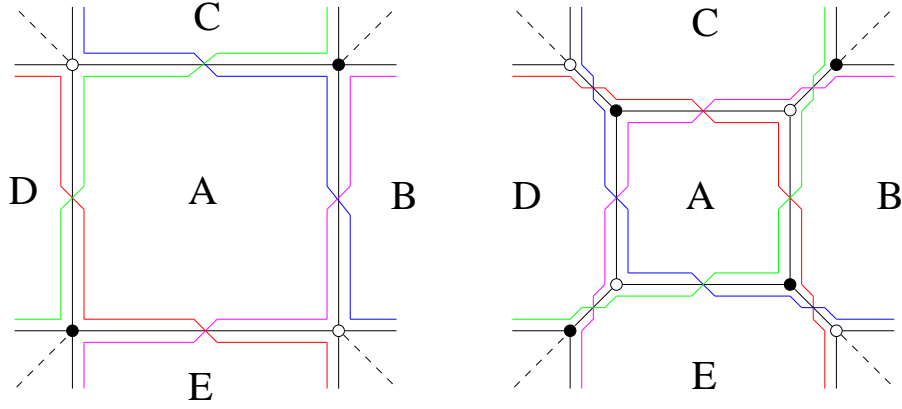}
\caption{Seiberg Duality on the zig-zag paths.}
\label{fig:SD}
\end{center}
\end{figure}

In this section we study the invariance of the formula (\ref{eq:generalzz})
under Seiberg duality.  
The duality on the dimer and on the zig-zag paths
is shown in figure \ref{fig:SD}.
The  zig-zag paths involved in the duality are the 
four represented in the picture, the red (R), green (G), blue (B) and
magenta (M).
In the \emph{electric} case the zig-zag paths that are involved 
in the duality are 
\begin{eqnarray}\label{eq:ZZele}
zz_R &=& X_{AE}\quad X_{DA}\quad \widetilde{zz}_R \nonumber \\
zz_G &=& X_{DA}\quad X_{AC}\quad \widetilde{zz}_G  \nonumber \\
zz_B &=& X_{AC}\quad X_{BA}\quad \widetilde{zz}_B  \nonumber \\
zz_M &=& X_{BA}\quad X_{AE}\quad \widetilde{zz}_M  
\end{eqnarray}
where $\widetilde{zz}_i$ is the part of the zig-zag part 
that does not transform
under the duality.
In the magnetic theory we have
\begin{eqnarray}\label{eq:ZZmagn}
zz'_R&=&Y_{DC}\quad Y_{CA}\quad Y_{AB}\quad Y_{BE}
 \quad \widetilde{zz}_R \nonumber \\
zz'_G&=&Y_{DE}\quad Y_{EA}\quad Y_{AB}\quad Y_{BC}\quad
\widetilde{zz}_G \nonumber \\
zz'_B&=&Y_{BE}\quad Y_{EA}\quad Y_{AD}\quad Y_{DC} \quad 
\widetilde{zz}_B \nonumber \\
zz'_M &=&Y_{BC}\quad Y_{CA}\quad Y_{AD}\quad Y_{DE}
\quad \widetilde{zz}_M 
\end{eqnarray}
The electric and magnetic $R$-charges are related by
\begin{eqnarray}\label{eq:SDrcharges}
&&
r^Y_{CA} =1 - r^X_{AC}, \quad  \quad r^Y_{DC} = r^X_{DA}+r^X_{AC} \nonumber \\
&&
r^Y_{AB} =1 - r^X_{BA}, \quad  \quad r^Y_{BE} = r^X_{BA}+r^X_{AE} \nonumber \\
&&
r^Y_{EA} =1 - r^X_{AE}, \quad  \quad r^Y_{BC} = r^X_{BA}+r^X_{AC} \nonumber \\
&&
r^Y_{AD} =1 - r^X_{DA}, \quad  \quad r^Y_{DE} = r^X_{DA}+r^X_{AE} 
\end{eqnarray}
It is know easy to check that index calculated in the electric phase
coincide with the one of the magnetic phase thanks to
(\ref{eq:SDrcharges}).

\section{Global symmetries}
\label{Sec:Glo}

In this section we show that the chemical potentials of the global
symmetries preserve the factorization of the \emph{off-shell} index
over the zig-zag paths.  There are two kind of global symmetries,
baryonic and flavor symmetries.  The first class of symmetries may be
visualized as a sub set of the $U(1)$ symmetries inside the $U(N)$
at each node.  The non anomalous baryonic symmetries are
obtained from the trace anomaly Tr$SU(N)^2_i U(1)_{B_j}$. This can be
visualized with the signed adjacency matrix. The kernel of this
operator defines the combinations of baryonic symmetries that decouple
in the IR or become anomalous.  The zig-zag paths
are uncharged under these symmetries, because they are gauge invariant
paths, or equivalently they are closed on the quiver.  This is
consistent with the expectation that the baryonic symmetries do not
contribute to the index.  On the other hand the flavor symmetries are
associated to the homologies of the paths in the tiling and they are
expected to contribute.  By assuming the factorization of the index
over the zig-zag paths
\begin{equation}\label{eq:factin}
\det M(t) = \prod_{i=1}^{Z}(1-t^{\sum_{j\in\{Z_i\}} (1-r_{j}^{(i)})})
\end{equation}
we  now prove that
\begin{equation}
\label{eq:factmu}
\det M(t) = \prod_{z=1}^{Z}
1-t^{\sum_{j\in \{Z_i\}} (1-r_{j}^{(i)})}
\mu_1^{-\sum_{j\in\{Z_i\}} F_j^{(i)}}
\mu_2^{-\sum_{j\in\{Z_i\}} \widetilde F_j^{(i)}})
\end{equation}
The index is a polynomial with three types of contributions
 $t^2$ and $t^{r_i}$ and $t^{2-r_i}$, where $r_i$
is the $R$-charge of the $i$-th scalar in the chiral multiplet.  Every term
in the polynomial is generically a set of disjoint closed loops in the
quiver, a gauge invariant string of bosonic and fermionic fields.
After adding the flavor symmetries the three possible contributions
change as
\begin{eqnarray} 
t^2 &\rightarrow&  t^2 \nonumber \\ 
t^{r_i}&\rightarrow&  t^{r_i} \mu_1^{F_i} \mu_{2}^{\widetilde F_i} \\
t^{2-r_i}&\rightarrow&  t^{2-r_i}  \mu_1^{-F_i} \mu_{2}^{-\widetilde F_i} \nonumber
\end{eqnarray}
By using the constraints from $NSVZ$ and the superpotential we can
convert the charge associated to a fermion $\psi_{ij}$ in the charge
associated to a product of bosons $\prod \phi_\alpha$, where $\alpha
\in I$ is a set of pairs of labels that parameterizes the fields
involved in this relation, we have
\begin{equation}
t^{2-r_{ij}} \mu_1^{-F_{ij}} \mu_2^{-\widetilde F_{ij}}
=
\prod_{\alpha \in I} t^{r_\alpha} \mu_1^{F_{\alpha}} \mu_2^{\widetilde F_{\alpha}}
\end{equation}
We can also convert the terms in the diagonal entries of $M(t)$, proportional
to $t^2$ in $t^{r_W}$ or $t^{r_F}$, where the exponent is the sum of
the charges of fields in a generic superpotential term or in a face in
the tiling. Putting everything together we observe that before
considering the flavor symmetries the index is a polynomial in
$P(t^{r_i})$ where $r_i$ represents the charge in the $i$-th scalar,
while after we add these symmetries the index is a polynomial in the
form $P \left( t^{r_i} \mu_1^{F_i} \mu_2^{\widetilde F_i} \right)$.  This shows
that the mesonic flavor symmetries preserve the factorization.

\section{Geometric formulation}
\label{Sec:Geo}
\begin{figure}
\begin{center}
\includegraphics[width=10cm]{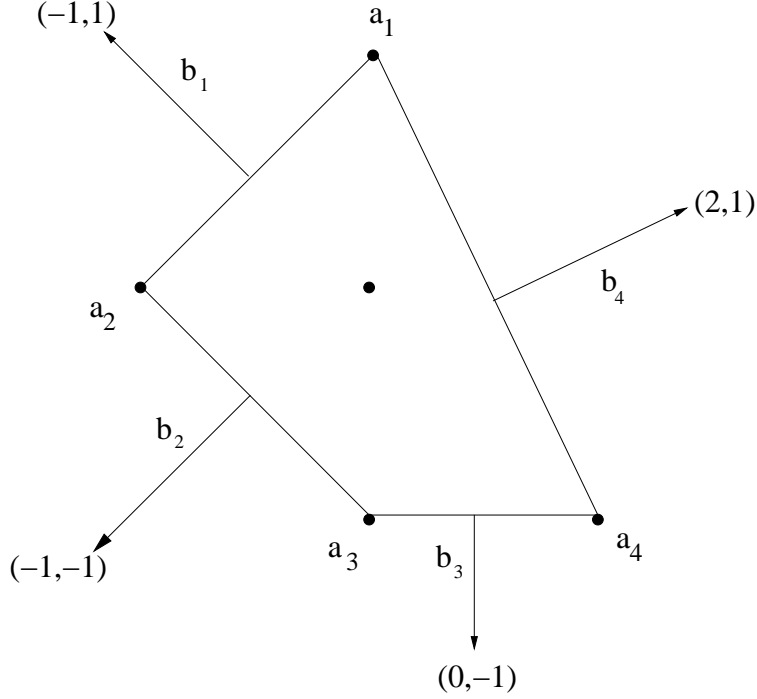}
\caption{Toric diagram of dP$_1$, primitive normals and charges.}
\label{fig:toricdp1}
\end{center}
\end{figure}

In this section we translate our formula of the index factorized over
the zig-zag in terms of toric geometry. 
As a standard procedure 
a set of variables $a_i$  is assigned 
to every external point of the toric diagram as in
\cite{Butti:2005vn} \footnote{In this case we restrict to the case
  without points on the edges.}.  They are constrained by $\sum
a_i=2$, which in the geometry represents the superpotential
constraint $R(W)=2$.  
A variable $b_i$ can be assigned to the primitive normals,
that are $1-1$ with the zig-zag paths, as
\begin{equation}
b_i=\sum_{j=1}^{i} a_i
\end{equation}
such that $b_d=2$ where $d$ is the number of external point of the
diagram.  
We give a pictorial representation of the toric diagram and 
the dual primitive vectors for dP$_1$ in figure \ref{fig:toricdp1}.
On the tiling $\pi b_i$ is the angle of intersection
of the zig-zag paths with the rombhi edges in the isoradial embedding
\cite{Hanany:2005ss}.
Every edges (fields) is crossed by two zig-zag paths and their $R$
charges are defined as
\begin{equation}
\left \{ 
\begin{array}{lcc}
R_{ij} = b_i-b_j&&i<j\\
R_{ij} = 2 -b_i+b_j&&i>j
\end{array}
\right.
\end{equation}
If more fields are crossed by the same pair of
paths they have the same charge.  
Once we obtained the formula for the
$R$-charges in terms of the geometry we can guess a 
formula that expresses the index in terms
of the $b_i$ variables.

A geometric formula that reproduces the
  field theory index is
\begin{equation}\label{eq:Indgeom0}
\det M_{geom}=
\prod_{i=1}^{d}(1-t^{\sum_{j} |\omega_{ij}| (1-R_{ij})})
\end{equation}
This formula holds in the minimal phase, where the number
of intersections between two zig-zag paths is fixed by 
\begin{equation}\label{eq:Indgeom}
\omega_{ij}=\langle \omega_i,\omega_j \rangle
=
\det
\left(
\begin{array}{cc}
p_i&q_i\\
p_j&q_j
\end{array}
\right)
\end{equation}
where $\omega_i = (p_i, q_i)$ are the primitive
normal vectors of the toric diagram.
After Seiberg duality one can end up with
non-minimal cases, where the number of intersections
is just bounded from below by 
$\langle \omega_i,\omega_j\rangle$.
In that case the formula is still valid because the
extra intersections always come in pairs 
with an opposite orientation and they cancel in 
(\ref{eq:Indgeom}) \cite{Hanany:2011bs}.

\subsection{dP$_1$}
\begin{figure}
\begin{center}
\includegraphics[width=10cm]{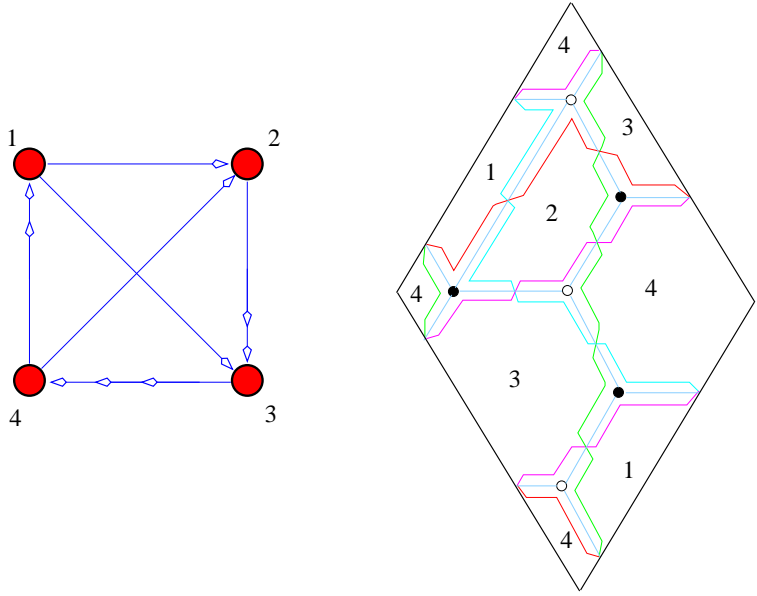}
\caption{Quiver, Tiling zig-zag paths and toric diagram of dP$_1$}
\label{fig:dp1}
\end{center}
\end{figure}
As an example we study the dP$_1$ model. The quiver the tiling and the toric
diagram are shown in figure \ref{fig:dp1}.
First we write the 
index in terms of the zig-zag paths, and than we use the geometric 
formula and show that the two formulas agree.
The superpotential is
\begin{equation}
W =
\epsilon_{\alpha \beta} X_{23}^{(\alpha)}X_{34}^{(\beta)}X_{42}
-
\epsilon_{\alpha \beta}  X_{12} X_{23}^{(\alpha)}X_{34}^{(3)}X_{41}^{(\beta)}
+
\epsilon_{\alpha \beta}X_{34}^{(\alpha)}X_{41}^{(\beta)}X_{13}
\end{equation}
The four perfect matchings related to the external points of the toric
diagram are
\begin{eqnarray}
&&
v_1=(0,1)~~ \rightarrow X_{13} X_{24} X_{34}^{(3)}\nonumber \\
&&
v_2=(-1,0)\rightarrow X_{23}^{(1)} X_{34}^{(1)}X_{41}^{(1)}  
\nonumber \\
&&
v_3=(0,-1) \rightarrow X_{12} X_{34}^{(1)} X_{34}^{(2)} 
\nonumber \\
&&
v_4=(1,-1) \rightarrow X_{34}^{(2)}  X_{23}^{(2)} X_{41}^{(2)}
\end{eqnarray}

\begin{eqnarray}
zz_1=X_{13}X_{34}^{(1)}X_{42}X_{23}^{(1)}X_{34}^{(3)}X_{41}^{(1)}
&&\quad\quad
zz_2=X_{41}^{(1)}X_{12}X_{23}^{(1)}X_{34}^{(2)}
\nonumber
\\
zz_3=X_{41}^{(2)}X_{12}X_{23}^{(2)}X_{34}^{(1)}
&&\quad\quad
zz_4=X_{34}^{(2)}X_{13}X_{41}^{(2)}X_{34}^{(3)}X_{23}^{(2)}X_{42}
\nonumber
\end{eqnarray}
The index is computed from the matrix
\begin{equation}
M(t)\!=\!
\left(
\begin{array}{cccc}
1\!-\!t^2 \!&\!-t^{r_{X_{12}}}\!&\!-t^{r_{X_{13}}} \!&\!
t^{2-r_{X_{41}^{(1)}}}\!+\!t^{2-r_{X_{41}^{(2)}}}
\!\\
t^{2-r_{X_{12}}}
\!&\!1\!-\!t^2 \!&\!-t^{r_{X_{23}^{(1)}}}\!-\!t^{r_{X_{23}^{(2)}}} \!&\!t^{2-r_{X_{42}}}\!\\
t^{2-r_{X_{13}}}\!&\!t^{2-r_{X_{23}^{(1)}}}\!+\!t^{2-r_{X_{23}^{(2)}}}\!&\!1\!-\!t^2 \!&\!
-t^{r_{X_{34}^{(1)}}}\!-\!t^{r_{X_{34}^{(2)}}}\!-\!t^{r_{X_{34}^{(3)}}}\!
\\ 
-t^{r_{X_{41}^{(1)}}}\!-\!t^{r_{X_{41}^{(2)}}}
\!&\!
-t^{r_{X_{42}}}\!&\!t^{2-r_{X_{34}^{(1)}}}\!+\!t^{2-r_{X_{34}^{(2)}}}\!+\!t^{2-r_{X_{34}^{(3)}}}
\!&\!1\!-\!t^2\!\\
\end{array}
\right)
\end{equation} 
The determinant of this matrix factorizes 
by imposing the
marginality  constraints and it is equivalent to
\begin{equation}
\label{Ppd1zz}
(1-t^{4-r_{zz_1}})(1-t^{6-r_{zz_2}})(1-t^{4-r_{zz_3}})(1-t^{6-r_{zz_4}})
\end{equation}
We now write the index from the geometric formula.
The $(p,q)$ web is parameterized by the four vectors
\begin{equation}
w_1=(-1,1) \quad,\quad
w_2=(-1,-1) \quad,\quad
w_3=(0,-1) \quad,\quad
w_4=(2,1)
\end{equation}
The $R$-charges of the fields intersecting on the zig-zag paths
can be written in terms of $b$ as
\begin{eqnarray} 
&&
R(1,2)=2 (b_2-b_1),\quad
R(1,3)=b_3-b_1,\quad
R(2,3)=b_3-b_2\nonumber \\
&&
R(2,4)=b_4-b_2,\quad
R(3,4)=2 (b_4-b_3),\quad
R(4,1)=3 (b_1-b_4+2)\nonumber
\end{eqnarray}
In terms of the $b$ variables the determinant is given by
(\ref{eq:Indgeom0}).  We have
\begin{equation}
\label{detb}
  \det M(t)=
  \left(1-t^{-3 b_1+b_2+2 b_3}\right) \left(1-t^{b_1+b_2-2 b_4+4}\right) 
  \left(1-t^{2 b_1-b_3-b_4+4}\right) \left(1-t^{-2 b_2-b_3+3 b_4}\right)
\end{equation}
The $b$ are related to
the $a$ variables as $b_i = \sum_{j=1}^{i} a_i$.
By assigning the $a_i$ variables to the external points
we can calculate the $R$-charge of the fields in terms of the $a_i$.
We  have 
\begin{equation}
\label{chargePM}
\begin{array}{c|c|c|c|c|c|c|c|c|c}
X_{12}&
X_{23}^{(1)}&
X_{23}^{(2)}&
X_{34}^{(1)}&
X_{34}^{(2)}&
X_{34}^{(3)}&
X_{41}^{(1)}&
X_{41}^{(2)}&
X_{13}&
X_{42}\\
\hline
a_3&a_2&a_4&a_2+a_3&a_3+a_4&a_1&a_2&a_4&a_1&a_1\\
\end{array}
\end{equation}
The expression in (\ref{detb}) coincides with (\ref{Ppd1zz}) after
substituting in the latter (\ref{chargePM}).

\section{Conclusions}
\label{Sec:Conc}

In this paper we observed that the superconformal index factorizes
over a set of gauge invariant paths on the dimer, called zig-zag
paths. 

We showed that this factorization remains valid 
also for theories with orbifold singularities, and without 
fixing the exact $R$-charge but on a generic set of $R_{trial}$ 
satisfying the marginality constraints. 

The zig-zag paths have an important role at geometrical level
because they give a mirror dual interpretation of the tiling.
Indeed, as observed in \cite{Feng:2005gw}, the zig-zag paths are both
 $(p, q)$ winding cycles in the dimer
and boundaries of the faces
in the tiling of the Riemann surface 
associated to a punctured region.
This allows a dual description in IIA in terms of mirror
D6 branes.
Our formulation in terms of the zig-zag paths
may be interesting for a \emph{mirror}
interpretation of the index.

A different duality, called specular duality, has been recently discovered
in \cite{Hanany:2012vc}. This duality exchanges the tiling with its mirror dual,
written in terms of the zig-zag paths.  Since the zig-zag paths have a
crucial role in the factorization of the index, it would be
interesting to analyze the relation among the indices in  specular dual
phases, as done here for the case of the usual Seiberg duality.

Another interesting development regards the relation with the
orientifolded theories. Indeed it is known that
the orientifold action 
on the tiling corresponds to a fixed line or fixed
point projection \cite{Franco:2007ii}. These projections are naturally
extended to the zig-zag paths. It would be nice to study the
relation between the zig-zag index and the orientifold in the tiling
and in the geometry.

A further line of investigation concerns the bipartite field theories
recently defined in
\cite{Franco:2012mm,Xie:2012mr,Heckman:2012jh,Franco:2012wv}.  Indeed,
even if they are not usually conformal, the zig-zag paths are well defined
on these theories.
It would be interesting to understand if the
formula we discussed in this paper has some field theoretical or
geometrical interpretation in those cases.

Finally, as discussed in the text, the zig-zag path are in one to one
correspondence with extremal BPS mesons.
In \cite{Benvenuti:2005ja} it has been shown that the extremal BPS
mesons correspond to massless geodesics of semiclassical strings moving 
in the internal geometry.
It would be intriguing to investigate 
possible connections  
between this hamiltonian system and the 
factorization of the superconformal index.

\section*{Acknowledgments}
We are grateful to A.~Zaffaroni for comments on the draft. We also
thank R.~Argurio, M.~Bianchi, S.~Franco and D.~Galloni for discussions
and comments.  P.A. is grateful to DOE-FG03-97ER40546 for fundings.
A.A. is grateful to the Institut de Physique Th\'eorique Philippe
Meyer at the \'Ecole Normale Sup\'erieure for fundings.
A.M. acknowledges funding by the Durham International Junior Research
Fellowship.

\appendix 
\section{Y$^{pq}$ theories}
\label{Ypqfam}
\begin{figure}
\begin{center}
\includegraphics[width=15cm]{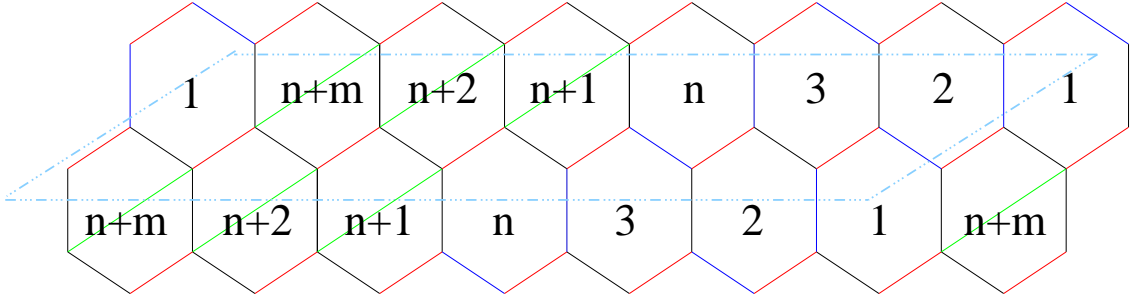}
\caption{Tiling for the Y$^{pq}$ theories. The different colors represent
the fields $U$ (black), $V$ (blue), $Y$ (red) and $Z$ (green). }
\label{fig:Ypq}
\end{center}
\end{figure}

In \cite{Gadde:2010en} the on shell superconformal index has been
computed for a generic $Y^{pq}$ theory \cite{Benvenuti:2004dy}, and
the authors guessed a generic formula by looking at different cases.
Here we show that by applying our formula in terms of the zig-zag
paths we can match their result on shell, but off shell the
factorization takes place over a different set of operators.  

A $Y^{pq}$ theory is a quiver gauge theory with $2p$ gauge groups.  In
figure \ref{fig:Ypq} we show the dimer and the four kind of fields
distinguished by their representation under the global symmetries.
From the figure one can extract the number of fields and their
charges.  They are given in the table
\begin{center}
\begin{tabular}{c||cc}
Field & Multiplicity & Charge \\
\hline
$Z$-green &$p-q$& $x$\\
$Y$-red&$p+q$& $y$ \\
$V$-blue&$2 q$ & $1+\frac{1}{2}(x-y)$\\
$U$-black&$2 p$& $1-\frac{1}{2}(x+y)$
\end{tabular}
\end{center}
The charges $x$ and $y$ are determined by 
$a$-maximization.
\begin{eqnarray}\label{exact}
x &=& \frac{(-4 p^2 - 2 p q + 3 q^2 + (2 p + q) \sqrt{4 p^2 - 3 q^2})}{(3 q^2)}
\nonumber \\
y &=& \frac{-4 p^2+2 p q+3 q^2+(2 p-q) \sqrt{4 p^2-3 q^2}}{3 q^2}
\end{eqnarray}
There are four kind of zig-zag paths. Two of them involve
all  the $Z$ and $p$($q$) $U$($V$) fields.
The other zig-zag paths exchange $Z$ with $Y$.
The contribution of these four paths to the index are
\begin{eqnarray}\label{mia}
&&
\sum_{j=1}^{Z_1}(1\!-\!r_{j}^{(1)})
=
\sum_{j=1}^{Z_2}(1\!-\!r_{j}^{(2)})
=2 p\!-\!((p\!-\!q)r_Z \!+\!  q r_V\! +\! p r_U\!)
=
\frac{(p\! -\! q) (2\! -\! x)\! +\! (p \!+ \!q) y}{2}
\nonumber \\
&&
\sum_{j=1}^{Z_3}(1\!-\!r_{j}^{(3)})
=
\sum_{j=1}^{Z_4}(1\!-\!r_{j}^{(4)})
= 
2(p\!+\!q)\!-\!(\!(p\!+\!q)r_Y \!+\!  q r_V \!+\! p r_U 
\!=\! 
\frac{(p \!+\! q) (2 \!-\! y) \!+\! (p \!-\! q) x}{2}
\nonumber
\end{eqnarray}
By comparing the formula obtained in \cite{Gadde:2010en} with our
formula we find that the two agree once the exact $R$-charge is
imposed.  If instead we just fix the constraints from the marginality
of the couplings, i.e. we keep $x$ and $y$ as generic variables
parameterizing a trial $R$-charge, we have
\begin{eqnarray} \label{rast} \det (M(t)) =
\prod_{i=1}^4
  (1-t^{\sum_{j=1}^{Z_i} (1-r_j^{(i)})}) 
\neq (1-t^{p (1 + (x - y)/2)})^2 (1-t^{p + 1/2 q
    (1 - 1/2 (x + y))})^2\nonumber \\
\end{eqnarray}
and the \emph{off-shell} index still factorizes over the zig-zag paths.

\end{document}